\documentclass[twocolumn,appendixfloats,tighten]{aastex6}
\usepackage{newtxtext,newtxmath}
\usepackage[T1]{fontenc}
\usepackage{ae,aecompl}

\usepackage{amsmath}	
\usepackage{amssymb}	

\usepackage{ctable}
\usepackage{url}
\usepackage{xspace}
\usepackage[normalem]{ulem}
\usepackage{hyperref}
\usepackage[all]{hypcap}
\usepackage{graphicx}

\def\msun{{\rm\,M_\odot}}

\def\msun{{\rm\,M_\odot}}

\newcommand{\be}{\begin{equation}}
\newcommand{\ee}{\end{equation}}

\newcommand{\fgas}{$ f_{\rm gas}$}
\newcommand{\rd}{$\rm R^*_{d}$}
\newcommand{\re}{$\rm R^*_{e}$}
\newcommand{\Msat}{ M_{\rm sat}}
\newcommand{\Mgas}{ M_{\rm gas}}
\newcommand{\Mstar}{ M_{\rm star}}
\newcommand{\rtide}{r_{\rm tidal}}
\def\h2{${\rm\,H_2}$}

\usepackage{color}

\begin{document}

\title{The Fate of Gas rich Satellites in Clusters}
\author{Mohammadtaher Safarzadeh  \&  Evan Scannapieco\\
School of Earth and Space Exploration, Arizona State University}
\email{Email: mts@asu.edu}



\begin{abstract}

We investigate the stellar mass loss of gas rich galaxies falling into clusters due to the change in the gravitational potential caused by the ram pressure stripping of their gas. We model the satellites with exponential stellar and gas disk profiles, assume rapid ram pressure stripping, and follow the stellar orbits in the shocked potential. Due to the change of the potential, the stars move from circular orbits to elliptical orbits with apocenters that are often outside the tidal radius, causing those stars to be stripped.  We explore the impact of the redshift of infall, gas fraction, satellite halo mass and cluster mass on this process. The puffing of the satellites makes them appear as ultra diffuse galaxies, and the stripped stars contribute to the intracluster light. Our results show these effects are most significant for less-massive satellites, which have larger gas fractions when they are accreted into clusters.  The preferential destruction of low mass systems causes the red fraction of cluster galaxies to be smaller at lower masses, an observation that is otherwise difficult to explain.

\end{abstract}

\section{Introduction}

Galaxies are observed to be gas rich at high redshifts \citep{Daddi:2010ee,Tacconi:2010cr} with large gas fractions \citep[\fgas $\equiv \rm \frac{M_{ gas}}{M_{gas}+M_{star}}$,][]{Geach:2011fr,Narayanan:2012bd,Popping:2012iv,MorokumaMatsui:2015gk,Popping:2015fs}. When galaxies fall into  clusters, their gas will be removed by ram pressure stripping \citep[RPS, ][]{Gunn:1972gx} in the hot halo of the cluster, an effect that has been well observed\citep{Kenney:2004fn,Chung:2007im,Boselli:2014cn,Brown:2017kf,Hayashi:2017cx} and modeled in numerical simulations \citep{Mayer:2006ee,Roediger:2007kr,Tonnesen:2009co,Steinhauser:2016da,Ruggiero:2017kc,Quilis:2017bc}. If the gas content of the system is removed either through RPS or other mechanisms, the stellar system will respond to the change in the gravitational potential and expand. Such puffing up of the stellar systems has been investigated  in relation to the size evolution of elliptical galaxies based on stellar winds or quasar feedback removing gas from the central part of the galaxy  \citep{Fan:2008bu,Damjanov:2009ik,Trujillo:2011jy,RagoneFigueroa:2011de}. 

Analytic estimates of expansion of the stellar system due to the change in the potential  have been provided by \citet{Biermann:1979fe} and \citet{Hills:1980fx} in the case of point source potentials. The change in the system depends on how fast  the mass loss takes place. If the mass loss timescale exceeds the dynamical time, the 
size of the system evolves as 
\be
\frac{R_f}{R_i}=\frac{1}{\epsilon},
\ee
where $\epsilon \equiv M_f/M_i$.
In case of rapid mass loss, on the other hand the final and initial radii of the system are related as
\be
\frac{R_f}{R_i}=\frac{\epsilon}{2\epsilon-1},
\ee
and the system becomes unbound for $\epsilon \le 0.5$. These analytic estimates, although insightful, do not capture the complexity of disk galaxies embedded in dark matter halos  \citep{RagoneFigueroa:2011de}. 

In this paper, we study the stellar orbits in gas rich dark matter halos that experience a change in the potential due to RPS of the gas in the intracluster medium. In this case, the dark matter, stars, and gas will each dominate the gravitational potential on different scales, and as the contribution from the gas vanishes the stars on initially circular orbits will experience a shock. As a consequence, the stellar orbits will expand and become more radial, and this puffing up of the stellar disk will lead to stars being stripped away due to tidal forces in the cluster.  We explore this processes over a large range of parameters including the satellite halo mass, redshift of infall, and the initial gas fraction. The stripped stellar mass will contribute to the intracluster light \citep[ICL, ][]{Murante:2004de,Zibetti:2005dc,Montes:2014bv,Contini:2014gi} and the puffing up of the stellar system will make the satellite appear as a diffuse system \citep{Dokkum:2015gy,Koda:2015kr,Roman:2017dv,Janssens:2017hp}.

In \S2 we present the method to calculate the new stellar orbits in the shocked potential. In \S3 we present our results for the fraction of stellar mass loss as a function of satellite and cluster properties. In  \S4 we suggest this mechanism can lead to the formation of ultra diffuse galaxies in the clusters, in \S5 we discuss the implications for the intracluster light, and in \S6 we discuss the impact on the red fraction of satellites. We summarize and discuss our results further in \S7.  Throughout this paper we adopt $\Lambda{\rm CDM}$ cosmology parameters based on Planck 2013 \citep{Collaboration:2014dt} $\Omega_{\rm m}=0.308$, $\Omega_{\rm b}=0.048$, $\Omega_{\Lambda}=0.693$, for the total matter, baryonic, and 
vacuum, densities, respectively, in units of the critical density,  $n=0.96$ is the tilt of the primordial power spectrum, and $\sigma_8=0.823$ is the rms value of the density fluctuations on the 8 $h^{-1}$ Mpc scale, where the Hubble constant is $H_0=100{h\,\rm km} \,{\rm s}^{-1}\,{\rm Mpc}^{-1}$ with $h=0.678$. 

\section{method}

We assume the stars are in initially circular orbits in a potential consisting of dark matter in an NFW profile  \citep{Navarro:1997if} together with the stellar disk and gas both in 
exponential profiles. The enclosed mass inside a given radius is 
\be
M_{\rm tot}(<r) = M_{\rm DM} (<r) + \Mstar (<r)  +\Mgas (<r).
\ee
The stellar system is modeled with an exponential profile:
\be
\Sigma^*=\Sigma^*_0 e^{-r/R^*_d},
\ee
where $R^*_d$ is scale length of the disk, which is computed as \re=1.67 \rd~where \re~is the half light (effective) radius of the disk. 
\re~is set to \re=0.015 $R_{200}(\Msat)$ \citep{Kravtsov:2013cy} and $\Sigma^*_0=\frac{M_*}{2\pi (R^*_d)^2}$ where stellar disk mass ($M_*)$ is assigned given a halo's redshift and mass following \citet[][hereafter B13]{Behroozi:2013fga}. 
We express the halo radius ($R_{200}$) defined as enclosing an overdensity of 200 times critical density of the universe at a given redshift, $\rho_{cr}(z)$,
so $M_{200}= (4\pi/3) 200 \rho_{cr}(z)R_{200}^3$.
The gas is modeled with a more extended exponential profile with $R_d^g=1.7 R^*_d$ \citep{Popping:2015fs} and  $\Sigma^g_0=\frac{M_{\rm gas}}{2\pi (R^g_d)^2}$ where $M_{\rm gas}=M_*$\fgas/(1-\fgas). We explore different ranges of \fgas in this paper. 

The enclosed mass of the dark matter halos is computed as
\be
M_{\rm DM} (<r) = M_{\rm sat} F(c) \frac{\ln (1+c x ) - c  r/x}{1+c r/x},
\ee
where $F(c)$ is defined as
\be
F(c)=\frac{1}{\ln(1+c)-c/(1+c)} , 
\ee
 $c$ is the concentration parameter of the halo, and $x= r/R_{200}(\Msat)$. 
The concentration parameter of the halos as a function of mass and redshift is estimated from \citet{Dutton:2014fm}. 
The adopted halo mass-size relation from \citet{Kravtsov:2013cy} is in excellent agreement with the 3D-HST+CANDELS observations of galaxies at $z<3$ \citep{vanderWel:2014hi}.

After accretion into the cluster, we assume the cold gas is removed by RPS on short timescales and therefore the system is left with only stars and dark matter to determine the new potential. The complete removal of gas content  is a good approximation for satellites on orbits with small impact parameters and more so for face on disks. \citet{Roediger:2007kr} modeled satellites with rotational velocities of 200 $\rm km/s$ moving the time varying intracluster medium (ICM) of two different clusters on various orbits and with various orientations.  All the gas was observed to be removed when the satellites were on small impact parameter orbits and more than 80\% of the gas was removed for satellites on large impact parameter orbits. In this study we assume our satellites become devoid of their cold gas disk which implies they have come close to the cluster core in their history. Note that the satellite's halo mass range in our study is below what is simulated  in \citet{Roediger:2007kr} and therefore have they lower gravitational restoring forces and consequently more prone to ram pressure stripping.

Initially the stellar orbits are assumed to be circular with velocities assigned as 
\be
V_c(r)=\left(\frac{G M_{\rm tot}(<r)}{r}\right)^{1/2}. 
\ee
The stellar orbits are shocked by the gas stripping and expand and exhibit a rosette like orbits. The time integration of the orbits is done with a $4^{\rm th}$ order Runge-Kutta method. Note that we do not model the puffing up of the dark matter halo or the impact of gas stripping in the outer radii and the drag that it exerts on the stellar disk in inner radii \citep{Smith:2012il}.

\begin{figure*}
\includegraphics[width=7in,height=2.5in]{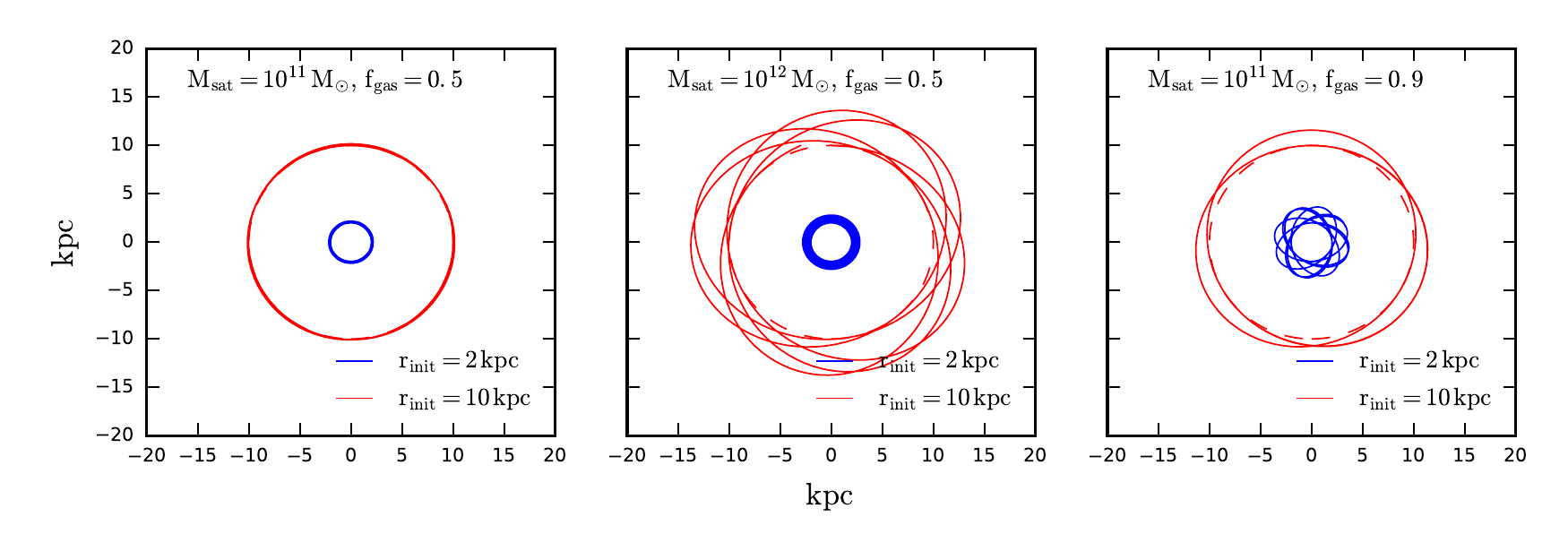}
\caption{The initial stellar orbits (dashed) and orbits after the gas is stripped away (solid). Stars initially in circular orbits in the potential of the halo exhibit rosette like orbits when the gas is stripped away leaving a shallower gravitational potential. On the left panel we show the case for two orbits initially at 2 and 10 kpc from the center in a halo with $ {M_h}=10^{11} \msun$, \fgas=0.5 and redshift ($z=1$). In the middle panel we show the result of gas stripping on the orbits of a $ {M_h}=10^{12} \msun$ with similar gas fractions. On the right panel we show orbits in $ {M_h}=10^{11} \msun$ halo with \fgas=0.9. The integration is carried out for 2 Gyr for all orbits shown. }
\label{f.example}
\end{figure*}

The left panel of Figure \ref{f.example} shows stellar orbits initially at two different radii (r=2,10 kpc) in a satellite with $ {M_h}=10^{11} \msun$ and initial \fgas=0.5. 
The dashed lines show the initial circular orbits and the solid lines show the trajectory of the orbits within 2 Gyr. In the middle panel we show the orbits in a satellite  with $ {M_h}=10^{12} \msun$ and \fgas=0.5. At higher halo masses, the impact of the gas stripping is more severe and the orbits expand to larger radii. In the right panel we 
show the orbits in a halo with $ {M_h}=10^{11} \msun$ but with \fgas=0.9. At higher gas fractions, and therefore larger contribution of gas to the initial potential of the orbits, the 
result of gas stripping becomes significantly larger. The redshift of the halos which determines the concentration parameter of the NFW potential is set to ($z=1$) for all cases.

\begin{figure}
\resizebox{3.5in}{!}{\includegraphics{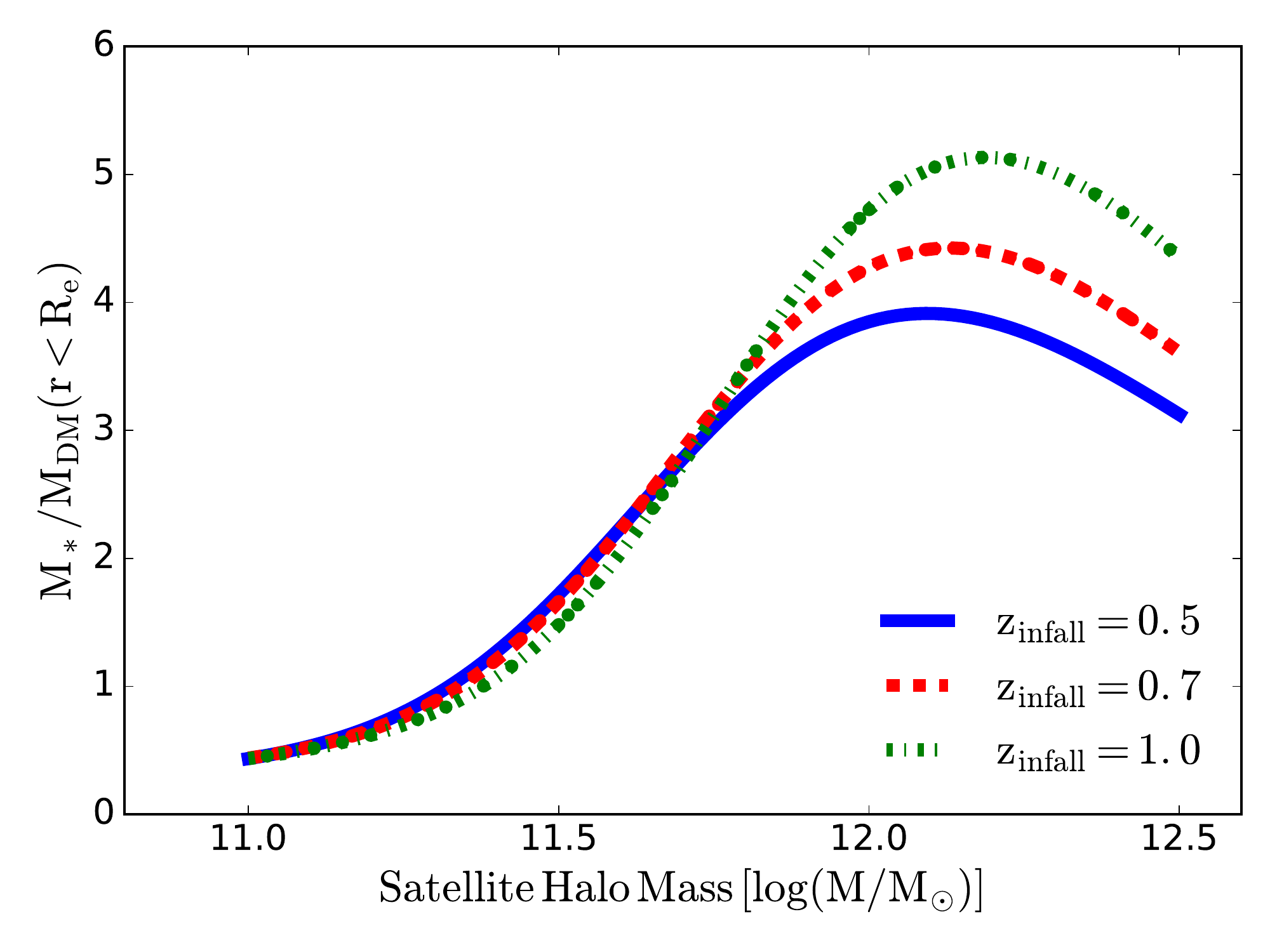}}
\caption{The ratio of the stellar mass to the dark matter mass inside the effective radius of the satellite. 
At higher halo masses, the baryonic disk dominates the potential as compared to the dark matter, and therefore gas stripping is more effective.
The baryon to dark matter relation is smaller in galaxies modeled in simulations such as EAGLE \citep{Schaller:2015cg}. However, we have assumed no feedback activity is present in the satellites when they are 
accreted into the clusters.
}
\label{f.ratio}
\end{figure}

In Figure \ref{f.ratio} we show the ratio of the stellar to dark matter mass inside the effective radius of the satellite. With increasing halo mass, the baryonic mass dominates the potential inside the effective radius and therefore the impact of gas stripping is stronger for more massive satellites \emph{at a fixed gas fraction}.
In order to determine what fraction of the stellar mass of the satellite is stripped away due to the puffing up process, 
we compare the maximal radial distance of the stars to the tidal radius of the satellite $\rtide$, computed as  \citep{Binney:2008wd}:
\be
\rtide= \left[ \frac{\Msat (<\rtide)}{3 M_{\rm cluster}(<D)}\right]^{1/3}  D,
\ee
where $D$ is the distance of the satellite from the center of the cluster. We find the root of the equation above using  Levenberg-Marquardt fitting method.

Although we have assumed the density structure is NFW for all halos, the impact of baryons on the initial dark matter halos is not negligible in that the central dark matter can undergo expansion in low-mass galaxies  \citep[e.g., ][]{Pontzen:2014hd}, while the dark matter likely contracts in more massive systems \citep[e.g., ][]{Duffy:2010ce,Dutton:2011dk}. 
We also do not consider the puffing up of the dark matter potential during gas stripping, which will enhance stellar mass loss as the dark matter potential drops at the center. 

On the other hand, our assumption of rapid stripping of gas may not hold for many of the orbits that satellites experience.
In such cases, the response of the stellar orbits to the change in the gravitational potential of the system would be different.  Investigating our theory in hydro-simulations of galaxy formation would self-consistently address these issues, and we plan to explore them in future studies.

\section{results}

\begin{figure}
\resizebox{3.5in}{!}{\includegraphics{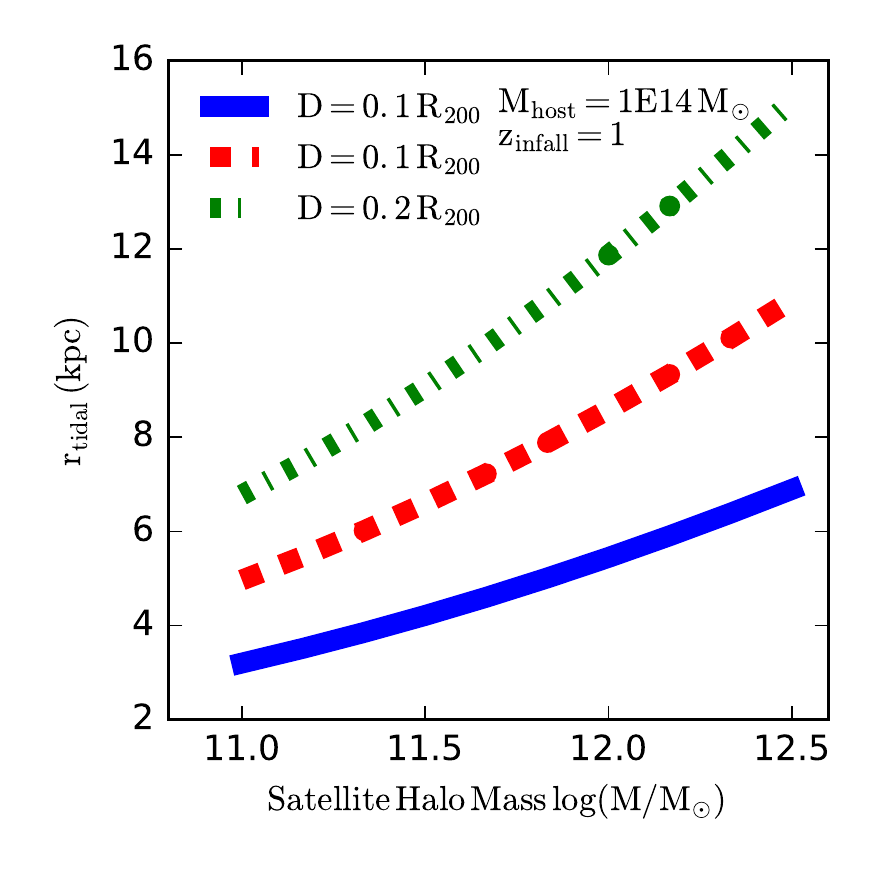}}
\resizebox{3.5in}{!}{\includegraphics{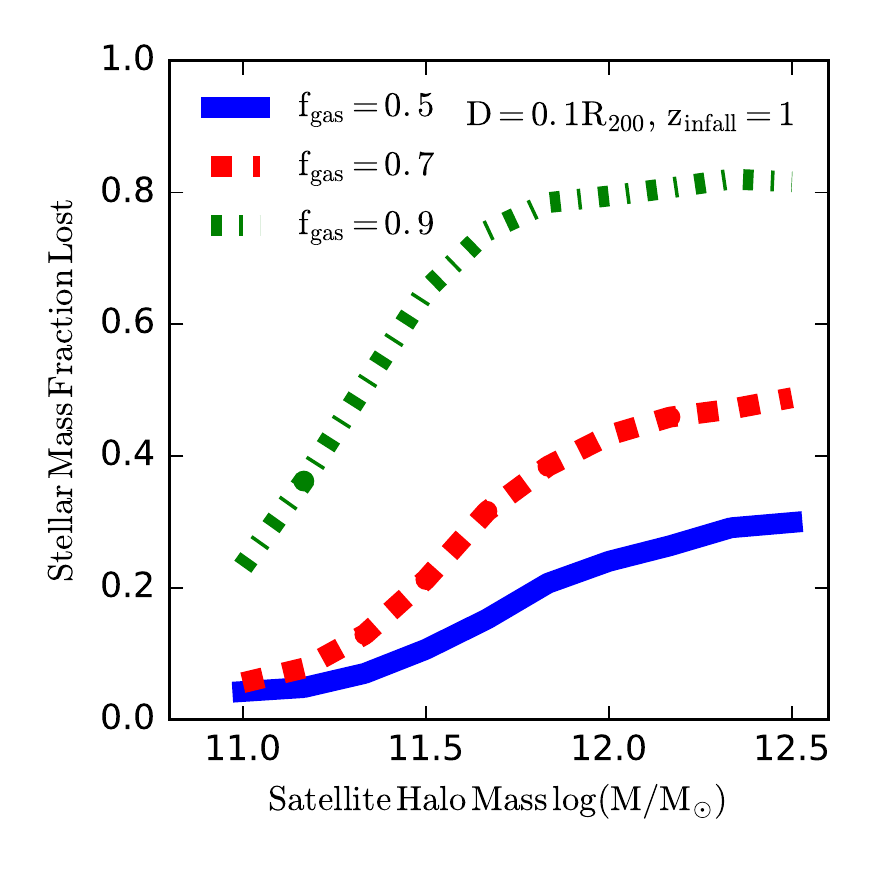}}
\caption{Top panel: Tidal radius as a function of satellite's halo mass for different cluster centric distances. Bottom panel: Fraction of the stellar mass that is lost due to the increase in orbital radii that occurs when the system loses its gas content.  We  show the effect of increasing the gas fraction of the satellite on the fraction of stellar mass that is stripped away for a satellite at $D=0.1R_{200}$ of the cluster center. The cluster mass is taken to be $10^{14}\msun$ with $R_{200}\approx$ 1 Mpc.}
\label{fgas_test}
\end{figure}

Top panel of Figure \ref{fgas_test} shows the Tidal radius as a function of satellite's halo mass for different cluster centric distances. Bottom panel of Figure \ref{fgas_test} shows the impact of $f_{\rm gas}$ on the fraction of stellar mass that is lost due to puffing up of the stellar disk as a function of satellite halo mass. On the left panel we show the result for three different gas fractions  (\fgas=0.5, 0.7 and 0.9) for satellites at a distance of 100 kpc from the center of a $10^{14} \msun$ cluster ($R_{200}$ for $10^{14}\msun \approx$ 1 Mpc). The stellar mass of the satellites is assigned according to B13, assuming the satellites are at $z=1$. The redshift affects both the stellar mass content and the concentration parameter of the satellite.  With a gas fraction of \fgas=0.9, more than 80\% of the stellar mass is stripped away for halos with $M_{h}>3\times10^{11}\msun$. The stellar mass loss is more severe for more massive halos despite the fact that their tidal radii are larger. 

\begin{figure*}
\centering
\includegraphics[width=7in,height=5.5in]{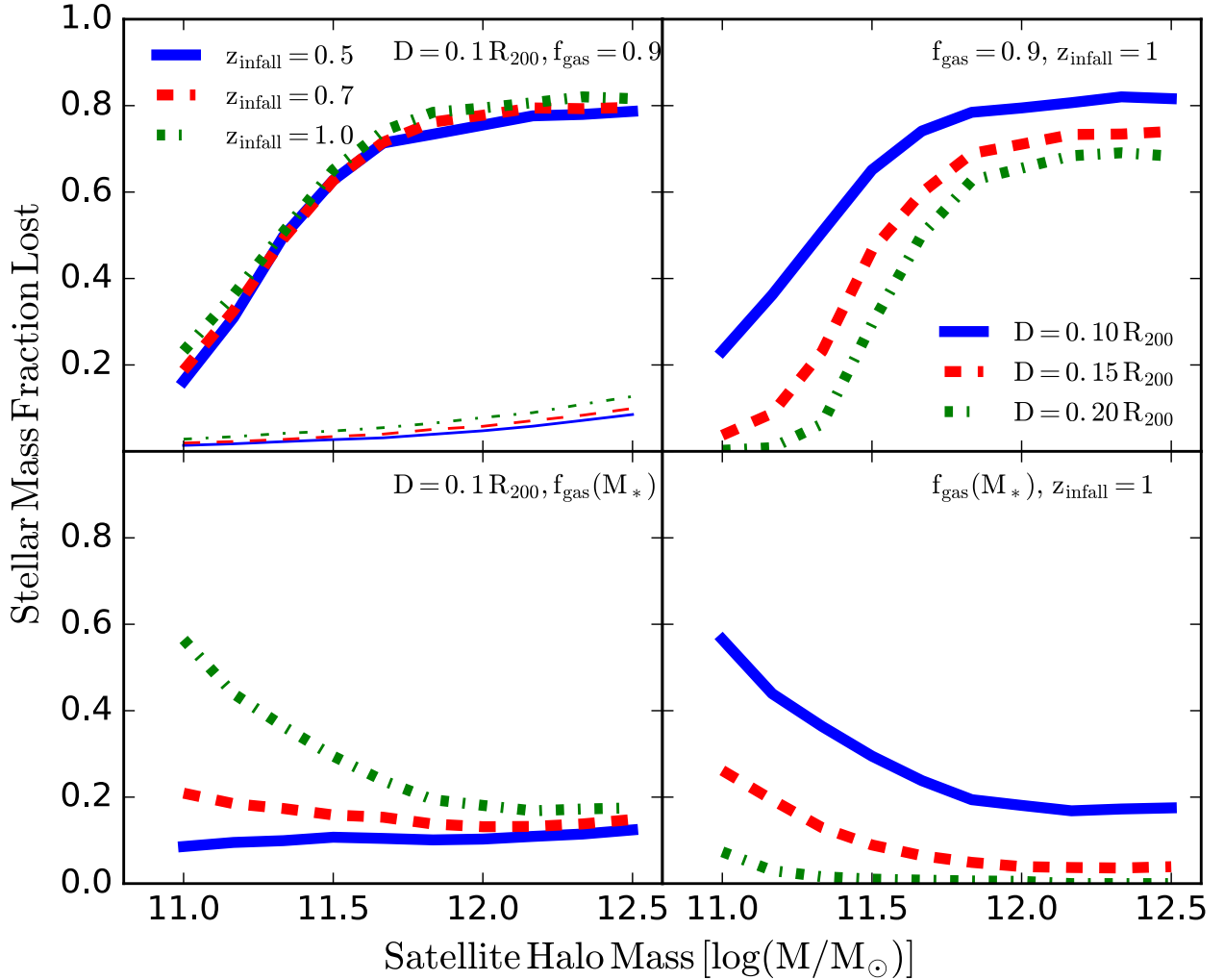}
\caption{Fraction of the stellar mass that is lost due to stellar disk puffing up after the system has lost its gas content as entering the host cluster. In the top-left panel, 
we show the trend with redshift of infall at a fixed gas fraction of 0.9 and cluster centric distance of $D=0.1 R_{200}$.The same is shown with thin lines for the case in which \fgas=0.0 is assigned to the satellites to illustrate the stripping effect without RPS.
The infall redshift at a fixed gas fraction only slightly impact the efficiency of the stripping. On the top-right panel we show the effect of cluster-centric distance at a fixed gas fraction of \fgas=0.9 and infall redshift set to $z=1$. In the bottom row, we show the case when the gas fractions are assigned following \citet{Popping:2012iv}. The cluster mass is taken to be $10^{14}\msun$ with $R_{200}\approx$ 1 Mpc.}
\label{zinfall_test}
\end{figure*}

The top-left panel of Figure \ref {zinfall_test} demonstrates the impact of the redshift of infall into the cluster at a fixed distance of 100 kpc from the center of a $10^{14}\msun$ cluster. 
The results are shown for \fgas=0.9 in thick lines and \fgas=0.0 in thin lines to illustrate the stripping effect without RPS. The infall redshift affects both the stellar mass content of the satellites that we assign following B13 and the concentration parameter of the satellites. There is a weak trend with infall redshift in that at higher redshifts the satellites tend to be more prone to tidal stripping. Satellites with $M_{h}>3\times10^{11}\msun$ lose about 80\% of their stellar mass regardless their accretion redshift. 

In the top right panel of Figure \ref {zinfall_test} we show the impact of cluster-centric distance  at given initial gas fraction for satellites with infall redshift set to $z=1$. The tidal radius becomes too large to strip away the stellar disk beyond $D=0.2 R_{200}$ of the cluster.

So far we have shown the results for a fixed gas fraction, however, both observations \citep{Popping:2012iv} and hydrodynamical simulations \citep{Narayanan:2012bd} indicate that cold gas fractions decrease with increasing stellar mass but increase with increasing redshift. In bottom row of Figure \ref {zinfall_test}, we show the results of assigning the gas fractions to galaxies following \citet{Popping:2012iv}\footnote[1]{we use the corrected formula presented in \citet{Popping:2015gf} } which estimated the gas fractions of galaxies in COSMOS field based on indirect measurement of the star formation rate densities as
\be
\label{eq:gas_frac_fit}
f_{\rm gas}=\frac{M_{\mathrm{gas}}}{M_{\mathrm{gas}} + M_*} = \frac{1}{\exp^{(\log{M_*} - A)/B} + 1},
\ee
where $A = 9 \times (1 + \frac{z}{0.017})^{0.03}$ and $B =1.1\times(1+z)^{-0.97}$. 

In this case, the trends are reversed from the constant gas fraction case presented in the top row of Figure \ref{zinfall_test}. The stellar mass fraction loss is more significant for less massive systems, due to their higher gas fractions. Furthermore, the redshift of infall has a significant effect. At $z=0.5$ the stellar mass fraction lost is almost at a constant level of 20\% for all satellite halo masses while at  higher redshifts we see less massive systems become more prone to stellar mass loss such that about 60\% of the stellar mass is lost in a satellite residing in a halo with mass $M_h=10^{11}\msun$ compared with 20\% for a $10^{12}\msun$ halo. 

In the bottom-right panel of Figure \ref{zinfall_test} we show the impact of cluster centric distance on the stripping efficiency. The stripping efficiency becomes negligible at distances beyond $\approx200$ kpc from the center while it sharply increases to high efficiencies close to $\approx 100$ kpc from the center, with the impact being more dramatic for less massive and more gas-rich systems. 

Lastly, we study the effect of cluster mass on satellite stellar mass loss at a fixed fraction of $R_{200}$. Galaxy clusters are observed out to redshifts $z\leq 2$ with masses in the range of $10^{14}-10^{15}\msun$ \citep{Vikhlinin:2009dt,McDonald:2013ea}. The highest redshift clusters discovered are  Cl 0218.3-0510 with a mass of $M_{200}\approx5.7\times10^{13}\msun$ at $z=1.62$ \citep{Tanaka:2010vj,Papovich:2010hm} and IDCS J1426.5+3508 with $M_{200}\approx5.3\times10^{14}\msun$ at $z=1.75$ \citep{Stanford:2012ua}. Recently \citet{Wang:2016in} discovered a galaxy cluster at $z=2.5$ with star forming galaxies in its core with mass of $M_{200}\approx8\times10^{13}\msun$. We only explore the mass range of $10^{14}-10^{15}\msun$ which is relevant for clusters at $z<1$.

\begin{figure}
\resizebox{3.5in}{!}{\includegraphics{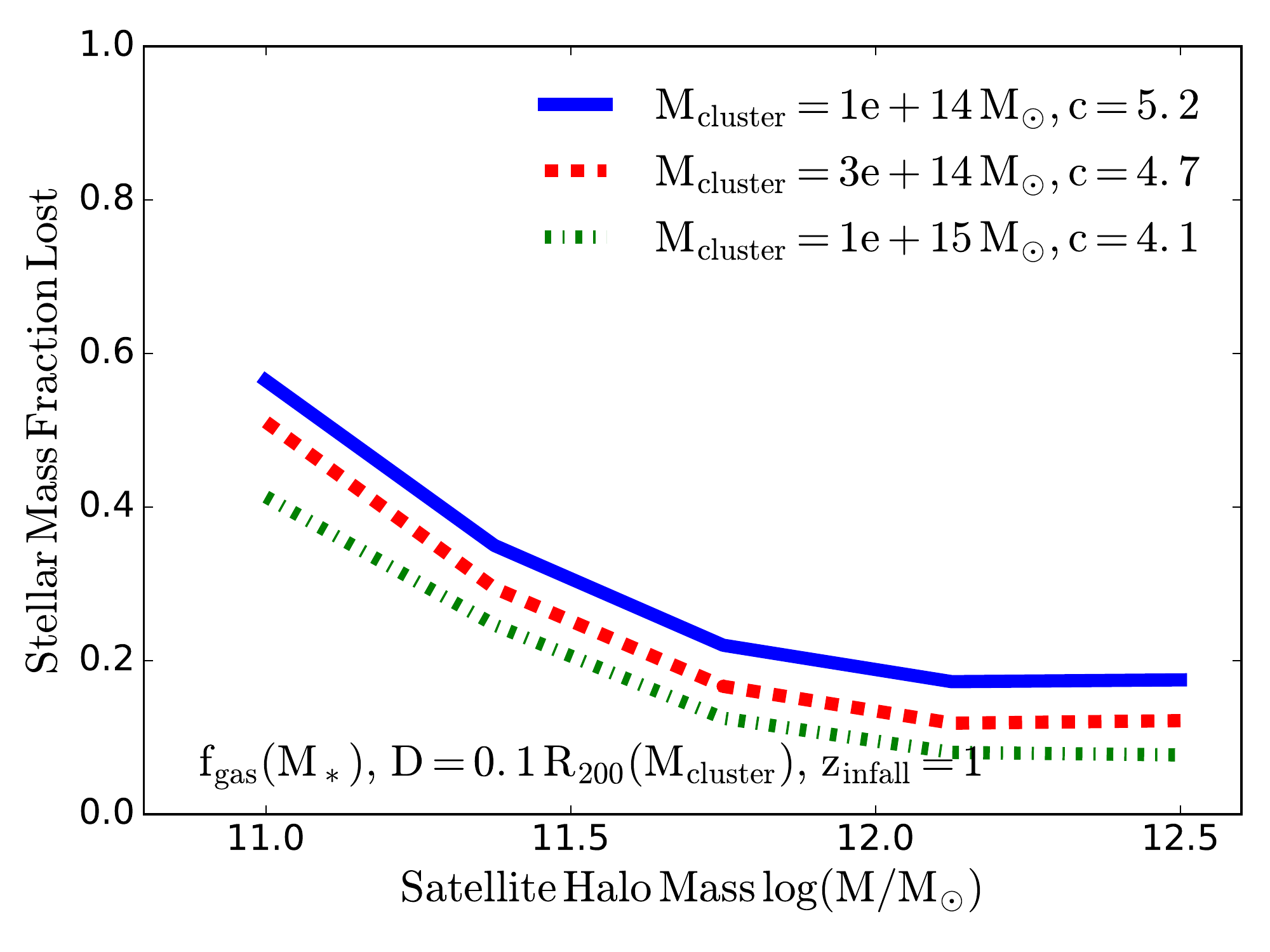}}
\caption{Fraction of the stellar mass that is lost due to stellar disk puffing up after the system has lost its gas content as entering the host cluster. This figure shows the trend at a fixed fraction of cluster's $D=0.1 \times R_{200}$ with cluster mass when the gas fractions are assigned based on \citet{Popping:2012iv} and stellar masses based on B13. The cluster mass and its concentration parameter is indicated in the legend. }
\label{f.cluster}
\end{figure}

Figure \ref{f.cluster} shows the result for three different cluster masses ($M_{\rm cluster}=10^{14},3\times10^{14},10^{15}$) at a constant distance of  $01 R_{200}(M_{\rm cluster})$. A higher fraction of the stellar mass of a satellite is lost for less massive halos, an effect that is related to the concentration parameter of the halo, which decreases for more massive hosts. 

\section{Puffed up satellites as Ultra Diffuse Galaxies}
The puffing up of the stellar disk due to RPS makes these systems appear diffuse.  Indeed diffuse systems are observed in galaxy clusters such as Coma  \citep{Dokkum:2015gy,Koda:2015kr,Yagi:2016cp} and Abell 2744 \citep{Janssens:2017hp} and are known as Ultra Diffuse Galaxies (UDGs). The stellar mass of UDGs is observed to be in the range of $\approx 10^7 -5\times10^8 \msun$. UDGs have effective radius of $r_e\approx1-5 $ kpc and effective surface brightness of $\mu_e(r)=25-28$ mag arcsec$^{-2}$ \citep{Koda:2015kr}  with essentially flat radial surface density distribution \citep{Janssens:2017hp}.

\begin{figure}
\resizebox{3.5in}{!}{\includegraphics{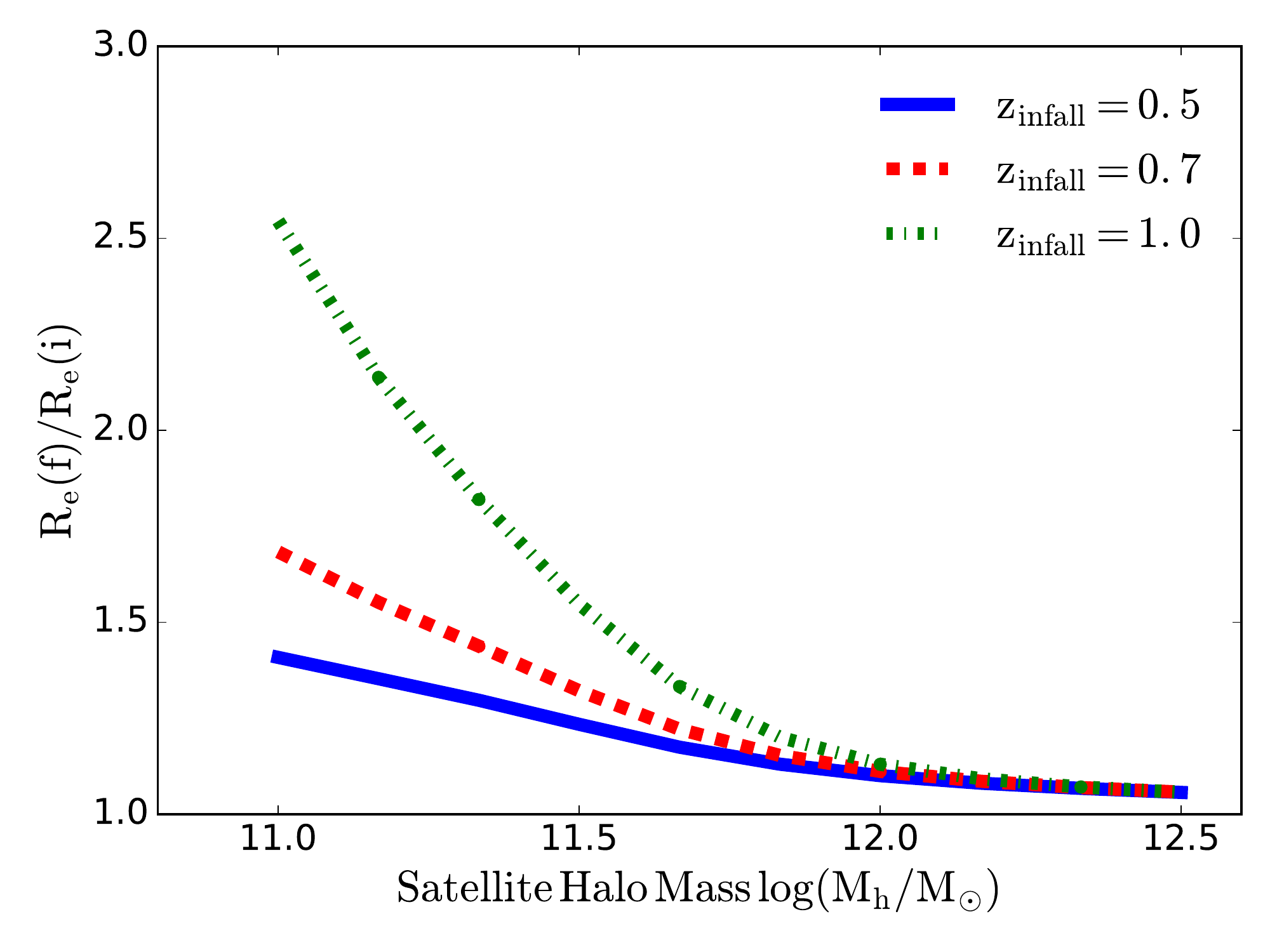}}
\caption{The ratio of the final to initial effective radius of the stellar disk in a satellite as a function of the satellite's halo mass. The gas fractions are assigned according to \citet{Popping:2012iv}. Higher expansion rate occurs for lower mass halos and at higher redshifts where both are because of higher gas fractions that is assigned for low mass systems and at higher redshifts.}
\label{f.R_e_ratio}
\end{figure}

In Figure \ref{f.R_e_ratio} we show the ratio of the effective radius, $R_e,$  of the puffed up system to the initial effective radius as a function of satellite halo mass.   In this figure, we have assigned the gas fractions following \citet{Popping:2012iv}, and
to compute $R_e,$ we have simply taken the time average radius of each star's orbit after the gas content is removed.  Here we see the impact is larger  for smaller halos because of the higher gas fraction that is assigned to them. The effect is also larger for higher redshifts, a fact that again is related to the increase of gas fraction with redshift. 

\begin{figure}
\resizebox{3.5in}{!}{\includegraphics{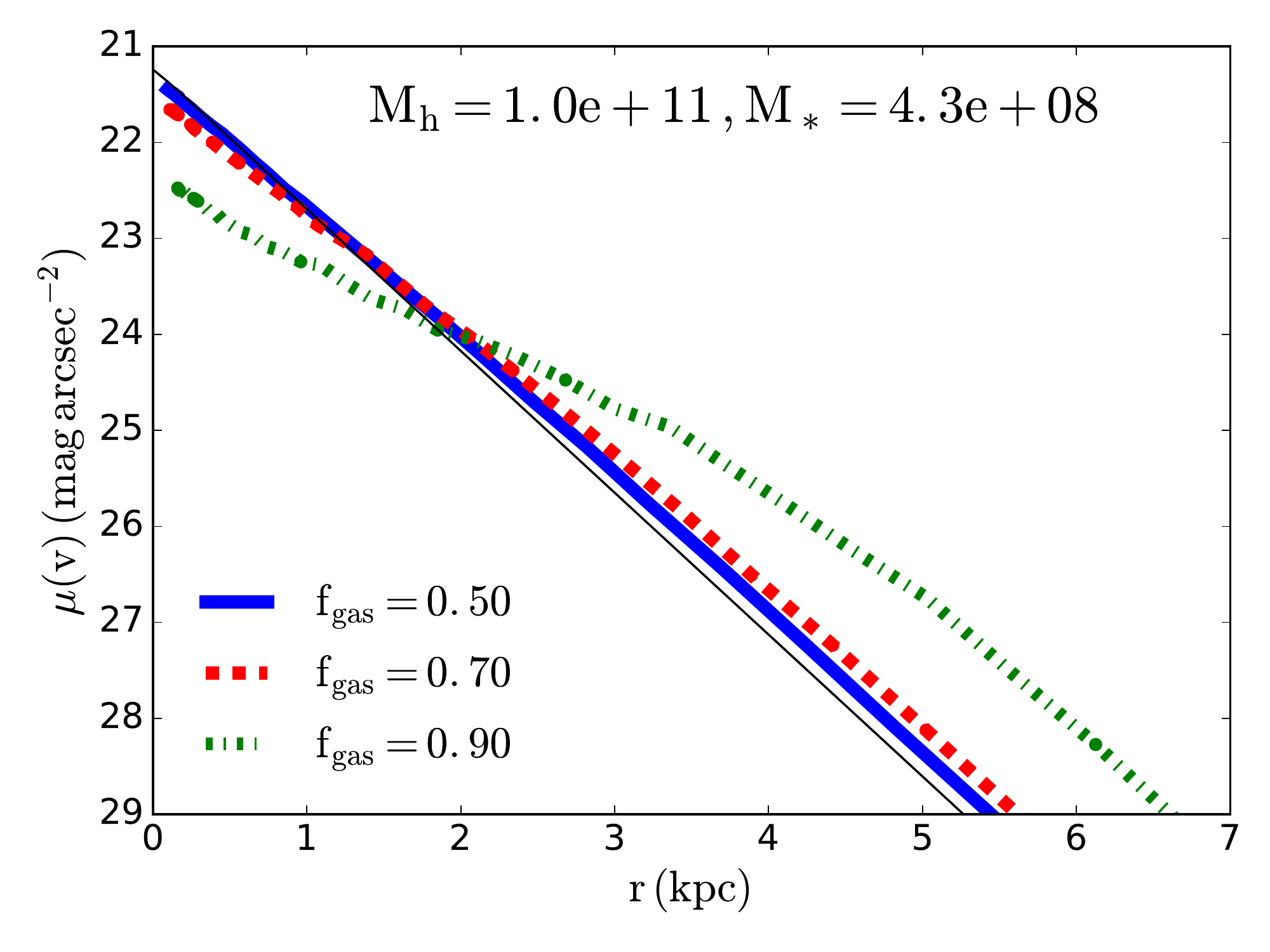}}
\caption{Impact of stellar disk puffing up on the surface brightness profile. The thin solid line shows the initial SB profile for a satellite with halo mass of $M_h=10^{11}\msun$ and stellar 
disk mass of $4.3\times10^8 \msun$ at $z=1$. The blue dashed(red dot-dashed, green dotted) line shows what the SB looks like if the gas fraction of the satellites at infall is \fgas=(0.5,0.7,09).
The gas fraction for a satellite with similar stellar mass at $z=1$ from \citet{Popping:2012iv} is \fgas=0.96, therefore observed satellites are closer to the green dotted line. }
\label{f.mu}
\end{figure}

As the stellar disks of the galaxies expand in reaction to the cold gas being stripped away, their surface brightness (SB) profiles  become shallower and more extended. Figure \ref{f.mu} shows how the SB evolves when the gas is stripped away from the system as a function of the initial gas fraction. We assume $z=0.5$ and the stellar mass is assigned based on B13 to a halo of $M_h=10^{11}\msun$. The SB starts to largely deviate from
its initial profile when the gas fraction of more than 0.9 is assumed for the satellites residing in halos of mass $M_h\approx10^{11}\msun$.

\begin{figure}
\resizebox{3.5in}{!}{\includegraphics{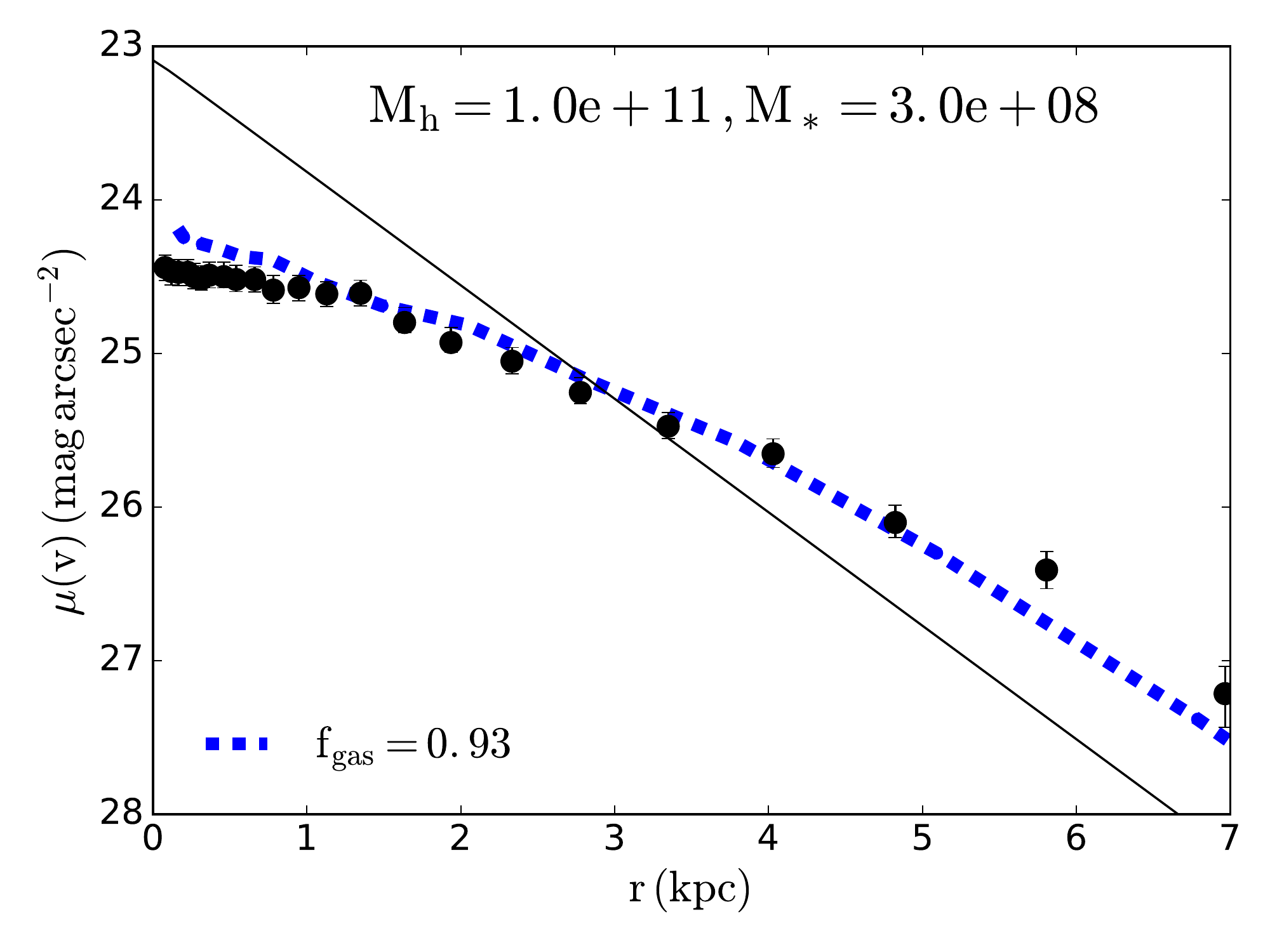}}
\caption{The change in the surface brightness profile of stellar disk with initial exponential profile when the gas content of the system is assumed to be ram pressure stripped away. The thin solid line shows the initial surface brightness profile and the thick solid line is the final surface brightness of the system.
The gas fraction before entering the cluster is assumed to be \fgas=0.93 and the stellar disk mass to be $M_*\approx 3\times10^{8}\msun$ in a halo mass of $M_h=10^{11} \msun$. We have set $R_e=0.03 R_{200}$ and $R_d^g=R_d^*$. The data points show the 
most diffuse galaxy DF44 observed in Coma cluster at $z=0.023$ \citep{Dokkum:2015gy}.}
\label{f.DF44}
\end{figure}

Figure \ref{f.DF44} shows the surface brightness profile of the initial and final state of the stellar disk after the gas is removed from a satellite of halo mass $M_h=10^{11} \msun$, initial gas fraction of \fgas=0.93 and stellar disk mass of $M_*\approx 3\times10^{8}\msun$. Here we adopt $R_e=0.03 R_{200}(\Msat)$ for the initial effective radius of the stellar system. 
The correlation factor is twice as adopted prior to this section but still within $2\sigma$ range of the relation found by \citet{Kravtsov:2013cy}. 
The result is compared to the observations of the most diffuse galaxy in the Coma cluster, DF44 \citep{Dokkum:2015gy} at $z=0.023$. The observed galaxy has an effective radius of 4.3 kpc. Here we see that with \fgas=0.93 we get a good fit to the data, suggesting that the observed diffuse galaxy could indeed be living in a halo of $M_h=10^{11} \msun$ at $z=0.5$. 
Even better fits to the data could be achieved with adopting higher initial gas fraction values.  Measurements of luminosity-weighted velocity dispersion of these systems could reveal what halo these systems reside in and help us constrain different theories regarding the UDGs.  

\section{Contribution to intracluster light}

Different methods have been devised to define and measure the diffuse intracluster light (ICL) observed in galaxy  clusters \citep{Rudick:2011eq}, leading to very different results. Tidal heating due to the interaction of $\approx L^*$ galaxies with substructure on the scale of 100 kpc was suggested by \citet{Gnedin:2003ei} to be the main mechanism unbinding the  stars making up of the ICL.  Using N-body simulations and implementing different approaches to model the stripping process in their semi-analytical model, \citet{Contini:2014gi} found that the main contribution to the ICL was from stripping of massive satellites with $M_*>10^{10.5}\msun $, though their ICL fraction depended on the resolution of the N-Body simulation.  On the other hand, \citet{Murante:2007jn} suggested that the merging events prior to the formation of the cluster are the main contributor to the ICL. Through a set of cosmological simulations, \citet{Martel:2012jp} concluded that destruction of galaxies by mergers dominates over the tidal stripping by about an order of magnitude at all redshifts. In these simulations, the bulk of the ICL was made by intermediate size satellites with halo mass in the range of $10^{11}-10^{12} \msun$.


Our results show that RPS has a larger impact on systems with higher gas fractions, and gas fractions continue to rise for less massive systems.
Therefore, sub $L^*$ systems with a halo mass of  $M_h\approx 10^{11} \msun$ or lower potentially contribute more to the ICL than previous thought.
We note that this conclusion is based on the assumption that all the satellites will experience the same  average orbit in the cluster, which is not the case as more massive systems are more prone to sink into the core of the cluster due to dynamical friction. The interplay between less resistive force to RPS in low mass systems and longer time to sink into the cluster core would need to be modeled in  more detail to make a definitive statement on what type of satellites are the main agents to make up the ICL light. 
However, to test the hypothesis we would need to have cosmological assembly histories and realistic orbits, which we are not modeling in our calculations here.


\section{red fraction of satellites in clusters}
The observed red fraction of satellites in clusters remains a challenge to reproduce in theoretical models \citep{Weinmann:2011fn,Luo:2016eo,Henriques:2017df}.  The red fraction of satellites is seen to increase as the satellite mass is increased at a fixed host halo mass \citep{Prescott:2011fh}. While the red fraction of field galaxies is larger for more massive systems, we expect the preferential destruction of gas rich satellites to play a significant role in galaxy clusters. Satellites being accreted into a cluster will become red as the supply of cold gas is terminated and the satellite has no fresh fuel to continue star formation. Because low mass systems have higher gas fractions and are more easily stripped,  this means that the low mass systems that are observed will be preferentially blue systems that have been recently accreted.  Thus the preferential destruction of satellites with lower masses will leave an imprint on the red fraction of satellites. 

In fact, there is observational evidence for such a mechanism. \citet{Hayashi:2017cx} detected gas rich satellites in the XMMXCS J2215.9-1738 cluster at $z = 1.46$ through CO (2-1) line emissions and the relative velocity of gas rich satellites versus cluster-centric distance points to them being accreted recently compared to gas poor satellites. Currently the models overproduce low mass passive satellites \citep{Luo:2016eo} and we anticipate incorporating disk puffing up in the models will bring them in better agreement with the observations of red  fraction of satellites, specifically for low mass systems. We note that our proposed mechanism is most efficient for those satellites reaching the inner parts of the cluster (i.e., $D<0.1 R_{200}$).
Therefore, the impact of this mechanism on the red fraction should be studied as a function of radii from the cluster.

\section{Summary and discussion}

We suggest a mechanism for enhancing the stripping of stars from galaxies in clusters due to the ram pressure stripping of their cold gas.
Initially circular orbits in a potential with gas, stars and dark matter, become rosette like when gas is removed, and this leads to more more radial orbits, some with apocenters reaching beyond the tidal radius of the galaxy. The efficiency of this mechanism largely depends on the initial gas fraction of the galaxies. 

When galaxy stellar masses are assigned based on abundance matching, and gas fractions are assigned based on stellar mass and redshift, we find that about 60\% of the stellar mass of a galaxy with a halo mass of $10^{11}\msun$ will be stripped away at a distance of $\approx 100$ kpc of a $10^{14}\msun$ cluster if the satellite enters the cluster at $z=1$.  
 At a fixed gas fraction, more massive galaxies are more prone to lose their stellar content by RPS of their cold gas, because they are more baryon dominated in their centers.  

However, because more massive  galaxies are also less gas rich, the efficiency drops at larger halo masses such that the fraction of the stellar mass lost reaches around 20\% for a galaxy residing in a halo with a mass of $10^{12}\msun$. Similarly, because they are more gas rich in general, less massive satellites are more prone to puff up and consequently lose their stellar content. Our results show that adopting realistic gas fractions results in strong stripping for those that approach radii of $D=0.1 R_{200}$ of the cluster. Therefore, only this subset of satellites is majorly affected by the proposed mechanism.

The stripped stars contribute to the intracluster light and when this mechanism is included theoretical models, it can potentially change the conclusions regarding the dominant sources of the ICL. The preferential destruction of low mass systems, can also potentially resolve the tension between the models and observations pointing to less massive satellites exhibiting lower red fractions.  As a galaxy puffs up, its surface brightness profile changes and it is possible that at the final stage the galaxy appears as an ultra diffuse galaxy depending on its initial gas fraction,  effective radius and halo mass. 

Several assumptions have gone into our work that future observations will better constrain. The ratio of the scale length of gas surface density to the stellar surface density is set to 1.7 based on the fits to the THINGS local sample of galaxies \citep{Leroy:2008jk}. The evolution of this ratio with redshift is not known but we note that lower values would lead to more puffing up of the stellar system as the gas will have a more compact profile.  

We do not model the star formation that plausibly continues for some time after the satellite is accreted into the cluster. The depletion timescale for galaxies given recent ALMA observations is around 1 Gyr \citep{Schinnerer:2016kz}. Although we assume no star formation occurs after the accretion, modeling this effect would be a task for hydrodynamical simulations \citep{Steinhauser:2016da} as attempting to capture the complicated physics of star formation and feedback effects on the gas distribution while the galaxy experiences RPS would be beyond the scope of this paper. 

\citet{Quilis:2017bc} studied the radial distribution of satellites of different masses in clusters through a set of hydrodynamical simulations and concluded satellites with stellar masses below $M_*\approx10^{10}\msun$ tend to be in the inner part of the cluster on radial orbits and more massive satellites tend to have eccentric orbits and reside on the external part of the cluster. They show that RPS  is more efficient for removing the cold gas of the satellites in the redshift range of $0.5<z<1$ where the fraction of unbounded gas soars for more massive satellites. 
Upcoming simulations that include clusters and resolve the stellar content of the satellites \citep[e.g., Illustris-TNG, ][]{Pillepich:2017vl} will be the testbed for the theory proposed here.  It remains to be seen how more detailed predictions will quantify how the puffing up of stellar disks due to RPS will impact the properties of ultradiffuse galaxies, the intracluster light,  and the red fraction of galaxies observed in cluster environments.

\acknowledgements

We would like to thank  Bruno Henriques, Yu Lu, and Gergo Popping for helpful discussions.  We thank the referee for their careful reading of this manuscript and many helpful comments. This work was supported by the National Science Foundation under grant AST14-07835 and by NASA under theory grant NNX15AK82G. 

\bibliographystyle{apj}
\bibliography{the_entire_lib}

\end{document}